\documentclass[prd,twocolumn,floatfix,amsmath,nofootinbib,amssymb,floatfix,preprintnumbers,showkeys]{revtex4}
\usepackage{graphicx,color,dcolumn,booktabs,bm,multirow}
\usepackage{longtable,lscape}
\usepackage{txfonts}
\usepackage{overpic}
\usepackage{amssymb}
\usepackage{array}
\usepackage{indentfirst}
\usepackage{feynmf}   
\usepackage{slashed}  
\usepackage{cases}
\usepackage{color}
\usepackage[T1]{fontenc}
\usepackage{multirow}
\usepackage{epstopdf}
\usepackage{tabularx}
\usepackage{graphicx,color,dcolumn,booktabs,bm}
\usepackage[colorlinks,
citecolor=blue,
anchorcolor=red,
menucolor=red,
linkcolor=red,
filecolor=red,
runcolor=red,
urlcolor=blue,
frenchlinks=red]{hyperref}
\usepackage{braket}
\usepackage{CJKutf8}
\usepackage{orcidlink}

\begin{document}
\begin{CJK}{UTF8}{gbsn}

\title{Study on the charmless decay modes of $Z_{cs}(3985)$}
\author{Yi-Zhuo Zhou$^{1}$}
\author{Shi-Dong Liu$^{4}$}
\author{Qi Wu$^{1}$}\email{wuqi@htu.edu.cn}
\author{Qing-Ping Ji$^{1}$}\email{jiqingping@htu.edu.cn}
\author{Dian-Yong Chen$^{2,3}$}\email{chendy@seu.edu.cn}
\author{Gang Li$^{4}$ }\email{gli@qfnu.edu.cn}
\affiliation{$^1$ School of Physics, Henan Normal University, Xinxiang 453007, China}
\affiliation{$^2$ School of Physics, Southeast University, Nanjing 210094, China}
\affiliation{$^3$ Lanzhou Center for Theoretical Physics, Lanzhou University, Lanzhou 730000, China}
\affiliation{$^4$ College of Physics and Engineering, Qufu Normal University, Qufu 273165, China}
\date{\today}

\begin{abstract}
In this work, we investigate the charmless decay modes of the hidden-charm tetraquark state with strangeness, $Z_{cs}(3985)$, assuming it to be an $S$-wave hadronic molecule of $D_{s}^{*-}D^{0}$ and $D_{s}^{-}D^{*0}$ with quantum numbers $I(J^{P}) = \frac{1}{2}(1^{+})$. Using an effective Lagrangian approach, we calculate the decay amplitudes for $Z_{cs} \to \mathcal{PV}$ (where $\mathcal P$ and $\mathcal V$ denote light pseudoscalar and vector mesons, respectively) via triangle loop diagrams. We use the experimental upper limit on $Z_c(3900)\to \pi\omega$ to constrain the cutoff parameter for both the dipole and tripole form factors. Numerical results from the dipole form factor show that the branching ratios for $Z_{cs} \to \pi K^*$ can reach up to the order of $10^{-2}$, which is of the same order of magnitude as the hidden charm decay modes and may be detected in future measurements. The branching ratios from tripole form factor are not consistent with those from the dipole form factor, indicating a certain degree of model dependence in the present work. The relative ratios of the branching ratios for different decay channels are found to be almost independent of the cutoff parameter. However, some ratios are highly dependent on the choice of the dipole or tripole form factor. The branching ratio of $Z_{cs} \to \eta K^{*}$ is found to be highly sensitive to the $\eta$-$\eta^\prime$ mixing angle, while that for $Z_{cs} \to \eta^\prime K^{*}$ is almost independent of the mixing angle, making it a clean channel for future experimental measurements.
\end{abstract}

\maketitle

\section{introduction}
\label{Sec:introduction}

The discovery of exotic hadronic states beyond the conventional quark model has profoundly reshaped our understanding of quantum chromodynamics (QCD) in the nonperturbative regime. Among the most compelling candidates for such exotic configurations are the charged charmoniumlike states, which contain a $c\bar{c}$ pair in combination with light quarks. Their unique quark content $c\bar{c}q\bar{q}$ makes them ideal probes for studying multiquark dynamics. Over the past two decades, a rich spectrum of such states has been uncovered, starting with $Z_c(4430)$ observed in the $\pi^-\psi(2S)$ invariant mass distribution in $B\to K\pi^-\psi(2S)$ decays by the Belle Collaboration in 2008~\cite{Belle:2007hrb,Belle:2013shl}. However, the \textit{BABAR} Collaboration searched for the $Z^-_c(4430)$ decay to $J/\psi\pi^-$ or $\psi(2S)\pi^-$ in $B^{-,0}\to J/\psi\pi^-K^{0,+}$ and $B^{-,0}\to \psi(2S)\pi^-K^{0,+}$ decays and no significant evidence was found for any of the processes investigated~\cite{BaBar:2008bxw}. In the same year that $Z_c(4430)$ was observed, the Belle Collaboration reported two additional $Z_c$ structures, at masses 4050 MeV/$c^2$ and 4250 MeV/$c^2$, in the $\pi^-\chi_{c1}$ system from $B\to K\pi^-\chi_{c1}$ decays~\cite{Belle:2008qeq}; however, a similar analysis by the \textit{BABAR} experiment did not confirm them~\cite{BaBar:2011hrz}.

The first confirmed charged charmoniumlike state is the $Z_c(3900)$ observed by the BESIII~\cite{BESIII:2013ris} and Belle~\cite{Belle:2013yex} Collaborations in the $\pi^\pm J/\psi$ invariant mass spectrum of $e^+ e^-\rightarrow \pi^+ \pi^- J/\psi$ at $\sqrt{s}=4.26$ GeV in 2013. It was confirmed by an analysis of CLEO-c data at $\sqrt{s}=4170$ MeV in the decay $\psi(4160)\to \pi^+\pi^-J/\psi$ and also found the neutral state $Z^0_c(3900)$~\cite{Xiao:2013iha}. A partner to the $Z_c(3900)$ with higher mass, named $Z_c(4020)$, was observed in the $\pi^\pm h_c$ mass spectrum of the process $e^+ e^- \to \pi^+\pi^-h_c$~\cite{BESIII:2013ouc,BESIII:2014gnk}. Moreover, the BESIII Collaborations also further confirmed them in other processes. The $Z_c(3900)$ and $Z_c(4020)$ were observed in the $(D\bar{D}^*)^\pm$ and $(D^*\bar{D}^*)^\pm$ invariant mass distribution of the processes $e^+ e^- \to(D\bar{D}^*)^\pm \pi^\mp$~\cite{BESIII:2013qmu} and $(D^*\bar{D}^*)^\pm \pi^\mp$~\cite{BESIII:2013mhi}, respectively. The evidence of the decay $Z^\pm_c(3900)\to \rho^\pm \eta_c$ was reported at $\sqrt{s}=4.226$ GeV with a significance of $3.9 \sigma$ and no significant signal was observed in a search for $Z^\pm_c(4020)\to \rho^\pm \eta_c$ with the same data samples~\cite{BESIII:2019rek}.

In 2014, the LHCb Collaboration confirmed the $Z_c(4430)$ in $B^0\to \psi^\prime \pi^- K^+$ and the spin-parity of this state is determined unambiguously to be $1^+$~\cite{LHCb:2014zfx}. This state, together with $Z_c(3900)$ and $Z_c(4020)$, provides irrefutable proof of the existence of charged charmoniumlike states and has been a topic of much debate. One possible clue for the interpretation of these $Z_c$ states is that their decay final states contain one charged pion and a charmonium state, such as $J/\psi$, $\psi(2S)$ and $h_c$. This feature indicates that $Z_c$ states most likely contain at least four constituent quarks and thereby are good candidates for tetraquark states~\cite{Maiani:2007wz,Ebert:2008kb,Patel:2014vua,Hadizadeh:2015cvx,Chen:2010ze,Qiao:2013dda,Wang:2013llv,Maiani:2013nmn,Deng:2015lca,Zhao:2014qva}. Another clue into the nature of these $Z_c$ states comes from the observation that they appear near the threshold composed of a pair of charmed and anticharmed mesons. Accordingly, the $Z_c(4430)$, $Z_c(3900)$ and $Z_c(4020)$ have been proposed as $\bar{D}_1 D^*$, $D\bar{D}^*$ and $D^*\bar{D}^*$ molecular states~\cite{Meng:2007fu,Rosner:2007mu,Liu:2007bf,Ding:2008mp,CLQCD:2009nvn,CLQCD:2010sna,Liu:2009qhy,Sun:2012zzd,Chen:2013omd,Gutsche:2014zda,Chen:2015igx,Li:2013xia,
Li:2014pfa,Xiao:2018kfx,Wu:2016ypc,Gong:2016hlt,Wu:2019vbk,Wu:2021udi,Wang:2022qxe,Wu:2023rrp,Wu:2025crk}, respectively. Besides the above exotic assignments, some kinematical effects have also been proposed to interpret these $Z_c$ states. For example, the $Z_c(4430)$ was proposed as arising from the $D^*\bar{D}_1$ threshold cusp effect in Ref.~\cite{Bugg:2008wu}; $Z_c(3900)$ and $Z_c(4020)$ can be reproduced with the initial single pion emission mechanism~\cite{Chen:2011xk,Chen:2013coa} or interpreted as a coupled channel cusp~\cite{Swanson:2014tra} or as resulting from triangle singularities~\cite{Szczepaniak:2015eza}.

Following the experimental confirmation of charmoniumlike tetraquark states, it is natural to expect the existence of hidden-charm tetraquark states containing strange quarks, based on SU(3) flavor symmetry. The hidden charm tetraquark states with strangeness have been predicted in various approaches, such as the relativistic quark model~\cite{Ebert:2008kb}, QCD sum rules~\cite{Lee:2008uy,Dias:2013qga}, and the single-kaon emission model~\cite{Chen:2013wca}. However, due to the low statistics of the data sample, the Belle Collaboration found that no clear structure was observed in the $K^\pm J/\psi$ system~\cite{Belle:2014fgf}. Until 2021, the hidden charm tetraquark states with strangeness, denoted as $Z_{cs}$, were found by the BESIII and LHCb Collaborations~\cite{BESIII:2020qkh,LHCb:2021uow}. The BESIII Collaboration observed a charged near-threshold structure $Z_{cs}(3985)$ in the $K^+$ recoil-mass spectra in $e^+ e^-\rightarrow K^+ (D^-_s D^{\ast0}+D^{\ast-}_s D^0)$ at the center-of-mass energy $\sqrt{s}=4.681$ GeV; and LHCb reported two resonances decaying into $K^\pm J/\psi$, namely $Z_{cs}(4000)$ and $Z_{cs}(4220)$. The resonance parameters of the $Z_{cs}(3985)$ and $Z_{cs}(4000)$ are shown in Table.~\ref{Tab:zcs}.

\begin{table}[htb]
\centering
\renewcommand{\arraystretch}{1.2}
    \caption{Resonance parameters of the newly reported $Z_{cs}$ states from different collaborations ~\cite{BESIII:2020qkh,LHCb:2021uow}. \label{Tab:zcs} }
    \setlength{\tabcolsep}{4.mm}
\begin{tabular}{ccc}
  \toprule[1pt]\toprule[1pt]
   Collaboration & Mass (MeV) & Width (MeV) \\
  \midrule[1pt]
  BES III~\cite{BESIII:2020qkh} & $3982.5^{+1.8}_{-2.6}\pm2.1$  & $12.8^{+5.3}_{-4.4}\pm3.0$  \\
  LHCb~\cite{LHCb:2021uow} &  $4003 \pm 6^{+4}_{-14}$  & $131 \pm 15 \pm 26$ \\
  \bottomrule[1pt]\bottomrule[1pt]
\end{tabular}
\end{table}

While the masses of the $Z_{cs}(3985)$ reported by BESIII and the $Z_{cs}(4000)$ reported by LHCb are similar, the two states exhibit significantly different widths.
The relationship between these two $Z_{cs}$ states remains under discussion~\cite{Yang:2020nrt,Meng:2021rdg,Ortega:2021enc,Maiani:2021tri,Shi:2021jyr,Giron:2021sla,Wang:2020rcx,Wang:2022clw,Karliner:2021qok,Han:2022fup,Wang:2022ckc}. The ratio of  branching fractions $\mathcal{B}(Z_{cs}(3985)^+\to K^+ J/\psi)/\mathcal{B}[Z_{cs}(3985)^+\to  (\bar{D}^0 D^{\ast+}_s+\bar{D}^{\ast0} D^+_s)]$ is determined to be less than $0.03$ at $90\%$ CL. by the BESIII Collaboration, which supports the $Z_{cs}(3985)$ and $Z_{cs}(4000)$ as two different states~\cite{BESIII:2023wqy}. The isospin partner of the $Z_{cs}(3985)^+$, $Z_{cs}(3985)^0$, was observed in $e^+ e^-\rightarrow K^0_S (D^-_s D^{\ast+}+D^{\ast-}_s D^+)$ with a significance of $4.6\sigma$ by the BESIII Collaboration~\cite{BESIII:2022qzr}. To deepen our understanding of these $Z_{cs}$ states and investigate the potential for new ones, the BESIII Collaboration searched for the $Z_{cs}$ in the decays $Z_{cs}\to K J/\psi$~\cite{BESIII:2023wqy} and $Z_{cs}\to K \psi(2S)$~\cite{BESIII:2024vjf}, but no significant $Z_{cs}$ states were observed.

The discovery of the $Z_{cs}(3985)$, $Z_{cs}(4000)$ and $Z_{cs}(4220)$ not only makes the spectrum of the charged charmonium-like states more complete, but also has motivated extensive discussions on their internal structures, such as hadronic molecules~\cite{Meng:2020ihj,Yang:2020nrt,Wang:2020rcx,Wang:2020htx,Dong:2020hxe,Chen:2020yvq,Chen:2021erj,Yan:2021tcp,Wu:2021ezz}, compact tetraquarks~\cite{Maiani:2021tri,Wan:2020oxt,Wang:2020iqt,Jin:2020yjn,Giron:2021sla}, or other phenomenological mechanisms~\cite{Wang:2020axi,Simonov:2020ozp,Ikeno:2020mra,Ge:2021sdq}, as well as their relations with $Z_c$ states. In Refs.~\cite{Yang:2020nrt,Meng:2020ihj,Maiani:2021tri}, the $Z_{cs}(3985)$ is identified as the strange partner of the $Z_c(3900)$. In Ref.~\cite{Chen:2021erj}, the authors systematically studied the $D^{(*)}\bar{D}^{(*)}$ and $D^{(*)}\bar{D}^{(*)}_s$ hadronic molecular states using the QCD sum rule method, and their results suggest that the $Z_{cs}(3985)$, $Z_{cs}(4000)$, and $Z_{cs}(4220)$ may be the possible strangeness partners of the $X(3872)$, $Z_c(3900)$ and $Z_c(4020)$, respectively. According to Refs.~\cite{Dong:2020hxe,Wang:2020dgr,Chen:2021uou,Shi:2021jyr,Wang:2020dmv,Du:2022jjv,Meng:2020ihj,Yang:2020nrt,Wang:2020rcx,Jin:2020yjn,Yan:2021tcp,Ortega:2021enc,
Han:2022fup,Ikeno:2021mcb,Wang:2022fdu}, there may exist heavier partners of the $Z_{cs}(3985)^-$, referred to as $Z^{\prime-}_{cs}$. This state is expected to predominantly decay into $D^{*-}_s D^{*0}$ with model-dependent masses falling between 4120 and 4200 MeV. An experimental measurement searching for $Z^{\prime-}_{cs}$ was performed by the BESIII Collaboration, and they found an excess of $Z^{\prime-}_{cs}\to D^{*-}_s D^{*0}(c.c.)$ candidates with a significance of $2.1\sigma$~\cite{BESIII:2022vxd}. The mass of $Z^{\prime-}_{cs}$ was measured to be $(4123.5\pm0.7_{stat.}\pm47_{syst.})$ MeV, and the width was not measured due to the small dataset.

In our previous work, we investigated the hidden charmed decays~\cite{Wu:2021ezz} and productions~\cite{Wu:2021cyc,Wu:2023fyh} of the $Z_{cs}$ states. Along with the decay behaviors of $Z_c(3900)$, as expected, the dominant decay mode of $Z_{cs}(3985)$ is the open charm channel, and the branching ratio of the hidden charm decay channels is about $20\%$. In addition to the open and hidden decay modes, exploring the possibility of other decay modes is helpful to enhance our understanding
of these $Z_{cs}$ states. Different from $Z_c$ and $Z_b$ states, in which the light quark and antiquark can annihilate into a photon and the remaining heavy quark and antiquark form a charmonium or bottomonium in the final state~\cite{Chen:2015igx,Cleven:2013sq}, the $Z_{cs}$ cannot undergo radiative decay via the same mechanism. From our previous work~\cite{Wu:2016ypc,Wang:2022qxe,Wu:2022hck}, the light hadron decay modes of the charmonium-/bottomoniumlike states are also not obviously negligible. In light of this, we investigate the charmless decay modes of the $Z_{cs}$ states in the present work.

This paper is organized as follows. After the Introduction, we present the model used in the estimation of the charmless decays of $Z_{cs}(3985)$. The numerical results and discussions are presented in  Sec.~\ref{Sec:Num}, and Sec.~\ref{sec:summary} is devoted to a short summary.

\section{Theoretical Framework}
\label{sec:formula}

\begin{figure*}[htb]
\begin{tabular}{ccc}
  \centering
 \includegraphics[width=4.2cm]{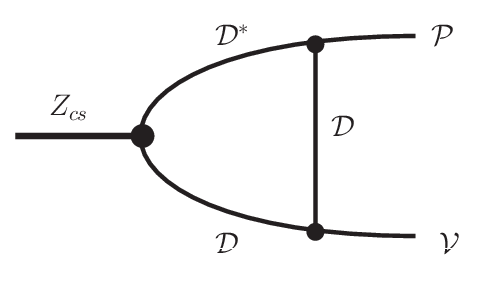}&
 \includegraphics[width=4.2cm]{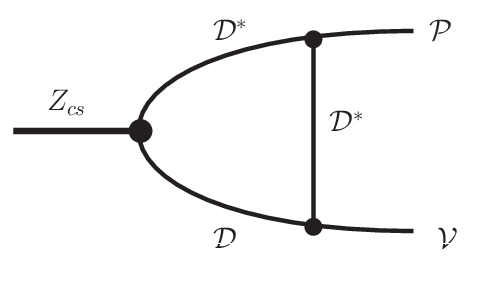}&
 \includegraphics[width=4.2cm]{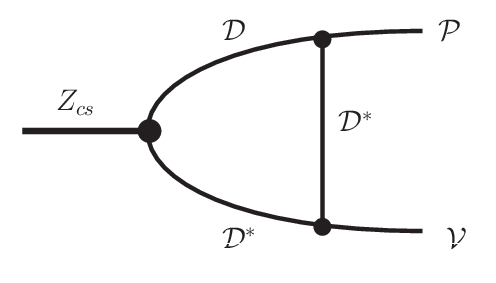} \\
 $(a)$ & $(b)$ & $(c)$ \\
 \end{tabular}
  \caption{Diagrams  contributing to $Z_{cs} \rightarrow \mathcal P \mathcal V$.}\label{Fig:Tri1}
\end{figure*}



\begin{table}[htp]
\centering
\renewcommand{\arraystretch}{1.2}
\caption{Loops (counterclockwise, starting from the top) contributing to $Z^-_{cs}\rightarrow \mathcal P\mathcal V$ with the final states $\bar{K}^0 \rho^-$, $K^- \rho^0/\omega$, $K^-\phi$, $\eta^{(\prime)}K^{*-}$, $\pi^0 K^{*-}$, and $\pi^- \bar{K}^{*0}$. \label{Table:loops}}
\setlength{\tabcolsep}{4mm}
\begin{tabular}{cc}
  \hline\hline
  Process & Loops \\
  \hline
  $Z^-_{cs}\rightarrow \bar{K}^0 \rho^-$ &
  $D^{*-}_s D^0 D^-$, $D^{*-}_s D^0 D^{*-}$, $D^{-}_s D^{*0} D^{*-}$ \\
  $Z^-_{cs}\rightarrow K^- \rho^0/\omega$ &
  $D^{*-}_s D^0 \bar{D}^0$, $D^{*-}_s D^0 \bar{D}^{*0}$, $D^{-}_s D^{*0} \bar{D}^{*0}$ \\
  $Z^-_{cs}\rightarrow K^- \phi$ &
  $D^{*0} D^-_s D^+_s$, $D^{*0} D^-_s D^{*+}_s$, $D^{0} D^{*-}_s D^{*+}_s$ \\
  \multirow{2}{*}{$Z^-_{cs}\rightarrow \eta^{(\prime)}K^{*-}$} &
  $D^{*-}_s D^0 D^{-}_s$, $D^{*-}_s D^0 D^{*-}_s$, $D^{-}_s D^{*0} D^{*-}_s$ \\
  &
  $D^{0} D^{*-}_s D^{*0}$, $D^{*0} D^{-}_s D^{0}$, $D^{*0} D^{-}_s D^{*0}$ \\
  $Z^-_{cs}\rightarrow \pi^0 K^{*-}$ &
  $D^{*0} D^{-}_s D^{0}$, $D^{*0} D^{-}_s D^{*0}$, $D^{0} D^{*-}_s D^{*0}$ \\
  $Z^-_{cs}\rightarrow \pi^- \bar{K}^{*0}$ &
  $D^{*0} D^{-}_s D^{+}$, $D^{*0} D^{-}_s D^{*+}$, $D^{0} D^{*-}_s D^{*+}$ \\
  \hline\hline
\end{tabular}
\end{table}

The state $Z^-_{cs}$ is assumed to be an $S$-wave hadronic molecule with quantum numbers $I(J^P)=\frac{1}{2}(1^+)$, formed as a superposition of the $D^{\ast-}_s D^0$ and $D^-_s D^{\ast0}$ configurations,
\begin{eqnarray}
|Z^-_{cs}\rangle &=& \frac {1} {\sqrt {2}} \left(|D^{\ast-}_s D^0\rangle+|D^-_s D^{\ast0}\rangle\right).
\label{eq:wavef}
\end{eqnarray}
Consequently, the coupling of $Z_{cs}$ to its constituent charmed mesons can be parametrized using the following effective Lagrangian:
\begin{eqnarray}
{\cal L}_{Z_{cs}}&=&\frac {g_{Z_{cs}}} {\sqrt {2}}Z^{\dagger\mu}_{cs}\left( \bar{D}^{\ast}_{s\mu} D + \bar{D}_s D^{\ast}_\mu \right ),
\label{eq:lagZcs}
\end{eqnarray}
where $g_{Z_{cs}}$ is the coupling constant.

Based on the heavy-quark limit and chiral symmetry, the Lagrangians describing interactions of light vector and pseudoscalar mesons take the form,
\begin{eqnarray}
{\cal L} &=& -ig_{D^{\ast }D
{\mathcal P}}\left( D^{\dag}_i \partial_\mu {\mathcal P}_{ij} D_j^{\ast \mu}-D_i^{\ast \mu\dagger} \partial_\mu {\mathcal P}_{ij}  D_j\right) \nonumber \\
&& +\frac{1}{2}g_{D^\ast D^\ast {\mathcal P}}\varepsilon _{\mu
\nu \alpha \beta }D_i^{\ast \mu \dag}\partial^\nu {\mathcal P}_{ij}  {\overset{
\leftrightarrow }{\partial }}{\!^{\alpha }} D_j^{\ast \beta } - ig_{{D}{D}\mathcal{V}} {D}_i^\dagger {\stackrel{\leftrightarrow}{\partial}}{\!_\mu} {D}^j(\mathcal{V}^\mu)^i_j \nonumber \\
&& -2f_{{D}^*{D}\mathcal{V}} \epsilon_{\mu\nu\alpha\beta}
(\partial^\mu \mathcal{V}^\nu)^i_j
({D}_i^\dagger{\stackrel{\leftrightarrow}{\partial}}{\!^\alpha} {D}^{*\beta j}-{D}_i^{*\beta\dagger}{\stackrel{\leftrightarrow}{\partial}}{\!^\alpha} {D}^j) \nonumber
\\
&&+ ig_{{D}^*{D}^*\mathcal{V}} {D}^{*\nu\dagger}_i {\stackrel{\leftrightarrow}{\partial}}{\!_\mu} {D}^{*j}_\nu(\mathcal{V}^\mu)^i_j \nonumber \\
&& +4if_{{D}^*{D}^*\mathcal{V}} {D}^{*\dagger}_{i\mu}(\partial^\mu \mathcal{V}^\nu-\partial^\nu
\mathcal{V}^\mu)^i_j {D}^{*j}_\nu +{\rm H.c.} , \label{eq:light-meson}
 \label{eq:LDDV}
 \end{eqnarray}
Here ${D}^{(\ast)\dagger}=(\bar{D}^{(\ast)0},D^{(\ast)-},D^{(\ast)-}_s)$ denotes the charmed-meson triplet, while $\mathcal P$ and ${\mathcal V}$ represent $3\times 3$ matrices for the octet pseudoscalar and nonet vector mesons, respectively
\begin{eqnarray}
    \label{eq-13}
    \mathcal{P} =
\left(\begin{array}{ccc}
\frac{\pi^{0}}{\sqrt 2}+ \frac{\beta\eta+\gamma\eta^\prime }{\sqrt{2}} &\pi^{+} &K^{+}\\
\pi^{-} &-\frac{\pi^{0}}{\sqrt 2}+ \frac{\beta\eta+\gamma\eta^\prime}{\sqrt{2}} &K^{0}\\
K^{-} &\bar K^{0} &-\gamma\eta+\beta\eta^\prime\end{array}\right),
\end{eqnarray}
\begin{eqnarray}
\mathcal{V} = \left(\begin{array}{ccc}\frac{\rho^0} {\sqrt {2}}+\frac {\omega} {\sqrt {2}}&\rho^+ & K^{*+} \\
\rho^- & -\frac {\rho^0} {\sqrt {2}} + \frac {\omega} {\sqrt {2}} & K^{*0} \\
K^{*-}& {\bar K}^{*0} & \phi \\
\end{array}\right) \, .
\end{eqnarray}
The physical $\eta$ and $\eta^\prime$ mesons are mixtures of the flavor eigenstates $n\bar{n}=(u\bar{u}+d\bar{d})/\sqrt{2}$ and $s\bar{s}$. Their wave functions are
\begin{eqnarray}
|\eta\rangle &=&\beta|n\bar{n}\rangle-\gamma|s\bar{s}\rangle,\nonumber \\
|\eta^\prime\rangle &=&\gamma|n\bar{n}\rangle+\beta|s\bar{s}\rangle \, , \label{eq:eta-etap mixing}
\end{eqnarray}
where $\beta=\cos{\alpha_P}$, $\gamma=\sin{\alpha_P}$, and $\alpha_P\simeq \theta_P+\arctan\sqrt{2}$. The empirical pseudoscalar mixing angle $\theta_P$ lies in the range $[-24.6^\circ , -11.5^\circ]$~\cite{ParticleDataGroup:2022pth}. In the present work, we adopt $\theta_P=-19.1^{\circ}$~\cite{MARK-III:1988crp,DM2:1988bfq}.

Using the effective Lagrangians above, the decay amplitudes for the triangle diagrams depicted in Fig.~\ref{Fig:Tri1} are straightforward to write down. For the processes $Z_{cs} (p_1) \to [D^{(*)} (q_1) {\bar D}^{(*)} (q_3)] D^{(*)} (q_2) \rightarrow \mathcal V (p_2) \mathcal P(p_3)$ illustrated in Fig.~\ref{Fig:Tri1}, the explicit amplitudes corresponding to the diagrams (a), (b), and (c) read
\begin{eqnarray}
\mathcal{M}_{a}&=&i^3 \int\frac{d^4 q_2}{(2\pi)^4}\Big[g_{Z_{cs}DD^*_s} \epsilon_{1\mu}\Big]\Big[g_{D^\ast D \mathcal P}p_{3\nu}\Big]
\Big[g_{DD\mathcal V}\epsilon_{2\alpha}^* (q^\alpha_3-q_2^\alpha)\Big]\nonumber\\
&&\times\frac{-g^{\mu\nu}+q^\mu_1 q^\nu_1 /m^2_1}{q^2_1-m^2_1}\frac{1}{q^2_2-m^2_2}\frac{1}{q_3^2-m_3^2} \mathcal{F} \left(q_2^{2}\right)\nonumber\,,\\
\mathcal{M}_{b}&=&i^3 \int\frac{d^4 q_2}{(2\pi)^4}\Big[g_{Z_{cs}DD^*_s} \epsilon_{1\kappa}\Big]\Big[\frac{1}{2}g_{D^\ast D^\ast\mathcal  P}\varepsilon_{\mu\nu\alpha\beta}p^\nu_3 (q^\alpha_1+q_2^\alpha)\Big]\nonumber\\
&&\times\Big[-2f_{D^\ast D\mathcal V}\varepsilon_{\rho\sigma\tau\xi}p^\rho_2 \epsilon^{*\sigma}_2 (q_2^\tau-q^\tau_3)\Big]\nonumber\\
&&\times\frac{-g^{\kappa\beta}+q^\kappa_1 q^\beta_1 /m^2_1}{q^2_1-m^2_1}\frac{-g^{\mu\xi}+q_2^{\mu}q_2^{\xi}/m^2_2}{q_2^2-m^2_2} \frac{1}{q^2_3-m^2_3} \mathcal{F} \left(q_2^{2}\right)\nonumber\,,\\
\mathcal{M}_{c}&=&i^3 \int\frac{d^4 q_2}{(2\pi)^4}\Big[g_{Z_{cs}DD^*_s} \epsilon_{1\rho} \Big]\Big[-g_{D^\ast D\mathcal P}p_{3\sigma}\Big]\nonumber\\
&&\times\Big[g_{D^\ast D^\ast \mathcal V}g_{\tau\theta} (q_{2\kappa}-q_{3\kappa})\epsilon^{*\kappa}_2
+4if_{D^\ast D^\ast \mathcal V}(-p_{2\tau} \epsilon_{2\theta}^*+p_{2\theta} \epsilon_{2\tau}^*)\Big]\nonumber\\
&&\times \frac{1}{q^2_1-m^2_1}\frac{-g^{\sigma\theta}+q_2^{\sigma} q_2^{\theta} /m^2_2}{q_2^2-m^2_2}\frac{-g^{\rho\tau}+q^{\rho}_3 q^{\tau}_3 /m^2_3}{q^2_3-m^2_3}\mathcal{F} \left(q_2^{2}\right).\nonumber\\
\end{eqnarray}
where $p_1 (\epsilon_1)$, $p_2 (\epsilon_2)$ and $p_3$ are the four-momenta
(polarization vectors) of the initial state $Z_{cs}$, final states $\mathcal{V}$  and $\mathcal{P}$, respectively. $q_1$, $q_3$, and $q_2$ are the four-momenta
of the up, down, and right charmed mesons in the triangle
loop, respectively.

In the present work, we adopt the following form factors for the exchanged intermediate meson, which is
\begin{eqnarray}
\mathcal{F}\left(q_2^{2}\right)=\Bigg(\frac{m^{2}-\Lambda^{2}}{q^{2}_2-\Lambda^{2}} \Bigg)^n\,,
\end{eqnarray}
where the parameter $\Lambda$ can be further reparameterized as $\Lambda_{D^{(\ast)}}=m_{D^{(\ast)}}+\alpha\Lambda_{\rm QCD} $ with $\Lambda_{\rm QCD}=0.22 \ {\rm GeV}$ and $m_{D^{(\ast)}}$ being the mass of the exchanged charmed meson. In practice, the value of $\alpha$ is usually determined by comparing theoretical estimates with the corresponding experimental measurements and its concrete value cannot be estimated from first principle. The introduced form factor is used to compensate the off-shell effects of the exchanged mesons partly as well as the structure effects of the involved mesons. In addition, it also plays the role of avoiding integration divergence.

\section{Numerical Results and discussions}
\label{Sec:Num}

\subsection{Coupling constants}

Before presenting the numerical results for $Z_{cs}\rightarrow \mathcal P\mathcal V$ decay, we first discuss the coupling constants involved in the present calculation. In the heavy quark and chiral limits, the couplings of charmed mesons to light vector mesons satisfy the following relations~\cite{Casalbuoni:1996pg,Cheng:2004ru},
\begin{eqnarray}
g_{{ D}{ D}\mathcal V} = g_{{ D}^*{ D}^*\mathcal V}=\frac{\beta g_V}{\sqrt{2}} , \quad f_{{ D}^*{ D}\mathcal V}=\frac{ f_{{ D}^*{ D}^*\mathcal V}}{m_{{ D}^*}}=\frac{\lambda g_V}{\sqrt{2}} \, , \nonumber\\
g_{{D}^{*} {D} \mathcal{P}}=\frac{2 g}{f_{\pi}} \sqrt{m_{{D}} m_{{D}^{*}}}, \quad g_{{D}^{*} {D}^{*} \mathcal{P}}=\frac{g_{{ D}^{*} { D} {\mathcal {P}}}}{\sqrt{m_{{D}} m_{{D}^{*}}}}. \label{eq:DDPorV}
\end{eqnarray}
Here, the parameter $g_V$ obeys $g_V = m_\rho/f_\pi$~\cite{Casalbuoni:1996pg}, with the pion decay constant $f_\pi = 132$ MeV and $\beta=0.9$. By matching the form factor from light-cone sum rules to lattice QCD results, one obtains $\lambda = 0.56$ GeV$^{-1}$ and $g = 0.59$~\cite{Isola:2003fh}.

Besides the coupling constants for charmed mesons and light mesons, the only coupling constant that remains to be determined is $g_{Z_{cs}}$. Unfortunately, the branching ratio of $Z_{cs}\to \bar{D}^0 D^{\ast+}_s+\bar{D}^{\ast0} D^+_s$ has not yet been measured by experiments. In Ref.~\cite{Wu:2021ezz}, the ratio of branching ratios $\mathcal{B}[Z_{cs}(3985)^+\to K^+ J/\psi]/\mathcal{B}[Z_{cs}(3985)^+\to  (\bar{D}^0 D^{\ast+}_s+\bar{D}^{\ast0} D^+_s)]$ is determined to be $0.01-0.1$ with the $g_{Z_{cs}}$ varying from 6.7 to 6.0 under the assumption that $D^*_s\bar{D}+\mathrm{H.c.}$, $J/\psi K$, and $\eta_c K^*$ are the three dominant decay channels of $Z_{cs}$. In the present work, we employ $g_{Z_{cs}}=6.5$ to estimate the decays $Z_{cs}\rightarrow \mathcal  P\mathcal V$, which corresponds to the upper limit $\mathcal{B}(Z_{cs}(3985)^+\to K^+ J/\psi)/\mathcal{B}[Z_{cs}(3985)^+\to  (\bar{D}^0 D^{\ast+}_s+\bar{D}^{\ast0} D^+_s)]<0.03$ measured by the BESIII Collaboration~\cite{BESIII:2023wqy}. Since the branching ratios of $Z_{cs}\rightarrow \mathcal P\mathcal V$ are proportional to $g^2_{Z_{cs}}$, the errors of the coupling constant have a non-negligible effect on the branching ratios. Given that there is no direct experimental measurement of the partial width of $Z_{cs}\to D^*_s \bar{D}$ so far, we set the uncertainty of $g_{Z_{cs}}$ to be $6.5^{+0.2}_{-0.5}$.

\subsection{$Z_c(3900)\to\mathcal{PV}$}

Considering that there is no experimental information on the charmless decays of $Z_{cs}$ and that the cutoff parameter $\alpha$ usually takes different values for different choices of $n$, one may infer useful information from the charmless decays of $Z_c(3900)$. In Ref.~\cite{BESIII:2015wge}, the decay $Z_c(3900)\to\pi\omega$ was searched for in the process  $e^+e^-\to Z_c(3900)\pi\to \omega\pi\pi$ by the BESIII Collaboration, and the decay width of $Z_c(3900)\to\pi\omega$ was determined to be smaller than $0.2\%$ of the total
width of $Z_c(3900)$ . In Table.~\ref{Tab:Zcbr}, we present the branching ratios of $Z_{c}(3900)\rightarrow \mathcal P\mathcal V$ for $n=2$ and $n=3$. We find that the experimental upper limit on the branching ratio of $Z_{c}(3900)\rightarrow \pi\omega$ can be reproduced with $\alpha=0.28$ for $n=2$ and $\alpha=2.23$ for $n=3$. In addition, the branching ratios of $Z_c(3900)\rightarrow \pi\omega$ for $n=2$ and $n=3$ are consistent with each other as we expected, while for $\eta^{(\prime)}\rho$, the branching ratios for $n=3$ are about 1--2 orders of magnitude larger that that for $n=2$. In what follows, we assume that the charmless decays of $Z_c(3900)$ and $Z_{cs}(3985)$ share the same $\alpha$ range based on SU(3)-flavor symmetry. We therefore use $\alpha=0.2-0.25$ and $\alpha=1.5-2$ to estimate the branching ratios of $Z_{cs}(3985)\rightarrow \mathcal P\mathcal V$ for $n=2$ and $n=3$, respectively.

\begin{table}[htb]
\centering
\setlength{\tabcolsep}{4pt}
\caption{Branching ratios of $Z_{c}(3900)\rightarrow \mathcal P\mathcal V$ for $n=2$ and $n=3$. The ranges for $n=2$ and $n=3$ correspond to $\alpha = 0.2- 0.25$ and $\alpha = 1.5- 2$, respectively.}  \label{Tab:Zcbr}
\begin{ruledtabular}
\begin{tabular}{lcc}
Channel & $n=2$   & $n=3$   \\
\hline \\[0.5ex]
$\pi\omega$ & $(0.5- 1.2)\times 10^{-3}$ & $(0.2- 1.1)\times 10^{-3}$   \\[2ex]
$\eta\rho$  & $(0.3- 0.8)\times 10^{-5}$ & $(1.6- 7.2)\times 10^{-4}$  \\[2ex]
$\eta^\prime\rho$  & $(0.1-0.3)\times 10^{-5}$ & $(0.8- 3.7)\times 10^{-4}$   \\[2ex]
\hline \\
Total  & $(0.5- 1.2)\times 10^{-3}$ & $(0.4- 2.2)\times 10^{-3}$ \\[0.5ex]
\end{tabular}
\end{ruledtabular}
\end{table}

\subsection{$Z_{cs}(3985)\to\mathcal{PV}$}

\begin{table}[htb]
\centering
\setlength{\tabcolsep}{4pt}
\caption{Branching ratios of $Z_{cs}(3985)\rightarrow \mathcal P\mathcal V$ for $n=2$ and $n=3$. The ranges for $n=2$ and $n=3$ correspond to $\alpha = 0.2- 0.25$ and $\alpha = 1.5- 2$, respectively. The errors are determined by the uncertainties of the coupling constants $g_{Z_{cs}}$  and the total width of $Z_{cs}(3985)$. }  \label{Tab:Zcsbr}
\begin{ruledtabular}
\begin{tabular}{lcc}
Channel~\footnote{Unless otherwise specified, the branching ratios $\mathcal{B}[Z_{cs}\to K\rho]$, $\mathcal{B}[Z_{cs}\to \pi K^*]$, etc., denote the sum over all decay channels for the charged $Z^-_{cs}$ state. Specifically, $Z^-_{cs}\to K^- \rho^0+\bar{K}^0 \rho^-$ for the $K\rho$ channel, and similarly for other channels.} & $n=2$   & $n=3$   \\
\hline \\[0.5ex]
$K\rho$ & $(1.7^{+0.7}_{-0.8}- 4.1^{+1.7}_{-2.1})\times 10^{-4}$ & $(6.9^{+2.9}_{-3.4}- 32^{+13}_{-16})\times 10^{-3}$   \\[2ex]
$K\omega$  & $(0.6^{+0.2}_{-0.3}- 1.4^{+0.6}_{-0.7})\times 10^{-4}$ & $(2.3^{+1.0}_{-1.1} - 11^{+4}_{-5})\times 10^{-3}$  \\[2ex]
$K\phi$  & $(0.8^{+0.3}_{-0.4}- 2.0^{+0.8}_{-1.0})\times 10^{-4}$ & $(4.0^{+1.7}_{-2.0} - 19^{+8}_{-9})\times 10^{-3}$   \\[2ex]
$\eta K^*$  & $(3.5^{+1.5}_{-1.8}- 8.6^{+3.6}_{-4.3})\times 10^{-7}$ & $(2.8^{+1.2}_{-1.4} - 12^{+5}_{-6})\times 10^{-6}$  \\[2ex]
$\eta^\prime K^*$  & $(1.1^{+0.5}_{-0.6}- 2.7^{+1.1}_{-1.4})\times 10^{-4}$ & $(6.6^{+2.8}_{-3.3} - 30^{+13}_{-15})\times 10^{-3}$   \\[2ex]
$\pi K^*$ & $(9.5^{+4.0}_{-4.8}- 24^{+10}_{-12})\times 10^{-3}$ & $(7.0^{+3.0}_{-3.5} - 32^{+14}_{-16})\times 10^{-3}$   \\[2ex]
\hline \\
Total  & $(9.9^{+4.1}_{-5.0}- 25^{+10}_{-13})\times 10^{-3}$ &  $(2.7^{+1.1}_{-1.4}- 12^{+5}_{-6})\times 10^{-2}$ \\[0.5ex]
\end{tabular}
\end{ruledtabular}
\end{table}

From Table.~\ref{Tab:Zcsbr}, it can be seen that the branching ratios of $Z_{cs}(3985)\rightarrow \pi K^*$ for $n=2$ and $n=3$ are consistent with each other within errors. For the other decay channels, the branching ratios for $n=3$ are one to two orders of magnitude larger than those for $n=2$, a behavior similar to the case of $Z_c(3900)$ and indicative of a certain degree of model dependence in the present work. In what follows, the discussion focuses mainly on $n=2$, since this case is sufficient to ensure the convergence of the integration. The dominant decay channel is $\pi K^*$, which is of the same order of magnitude as the hidden charm decays of $Z_{cs}(3985)$~\cite{Wu:2021ezz}. The branching ratios of $K\rho/\omega/\phi$ and $\eta^\prime K^*$ are of the order of $10^{-5}-10^{-4}$, which are two orders of magnitude lower than those
of the hidden charm decay channels. The smallest decay channel is $\eta K^*$, with a branching ratio of the order of $10^{-7}$.

\begin{figure}[!htb]
  \centering
 \includegraphics[width=8cm]{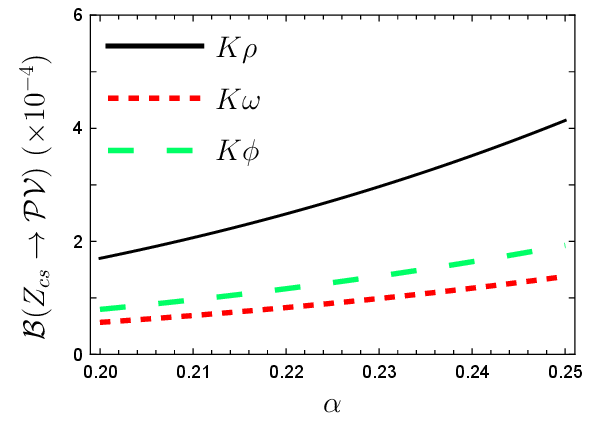}
  \caption{Branching ratios of $Z_{cs}\rightarrow \mathcal P\mathcal V$ in terms of $\alpha$ with $n=2$, where $\mathcal V=\rho, \omega, \phi$ and $\mathcal P=K$.}\label{Fig:pv1}
\end{figure}

\begin{figure}[!htb]
  \centering
 \includegraphics[width=8cm]{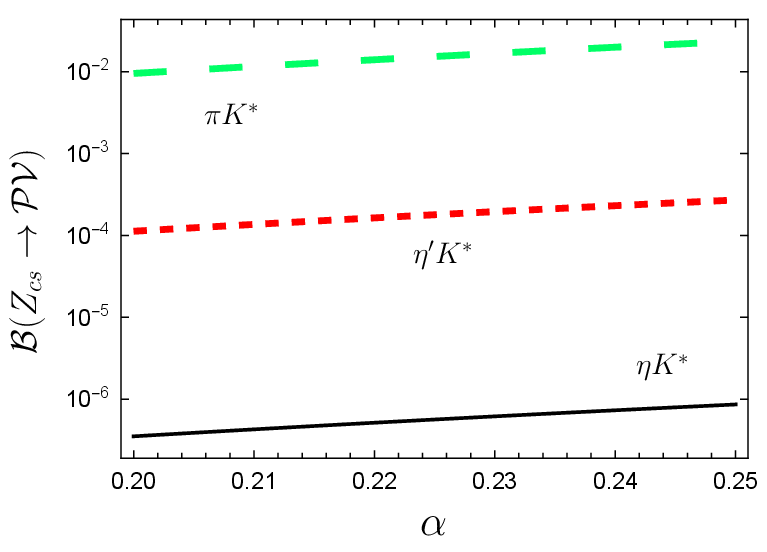}
  \caption{Branching ratios of $Z_{cs}\rightarrow \mathcal P\mathcal V$ in terms of $\alpha$ with $n=2$, where $\mathcal V=K^*$ and $\mathcal P=\eta, \eta^\prime, \pi$.}\label{Fig:pv2}
\end{figure}

In Figs.~\ref{Fig:pv1} and \ref{Fig:pv2}, we present the $\alpha$ dependence of the branching ratios of $Z_{cs}\rightarrow K\rho/\omega/\phi$ and $Z_{cs}\rightarrow \eta^{(\prime)}/\pi K^*$, respectively, with the $\eta-\eta^\prime$ mixing angle fixed at $\theta_P=-19.1^\circ$. For $\alpha$ in the range of $0.2$ to $0.25$, the $\alpha$ dependence of the branching ratios is not too strong.

\begin{figure}[!htb]
  \centering
 \includegraphics[width=8.5cm]{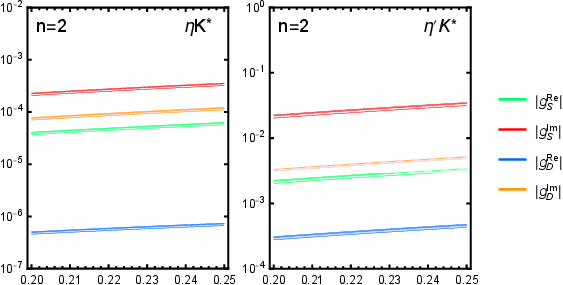}
  \caption{$\alpha$ dependence of $|g_D^{\rm Re}|$, $|g_D^{\rm Im}|$, $|g_S^{\rm Re}|$, and $|g_S^{\rm Im}|$ with $n=2$. The shaded areas result from the errors of the coupling constants.}\label{Fig:etaKstar_n2}
\end{figure}

\begin{figure}[!htb]
  \centering
 \includegraphics[width=8.5cm]{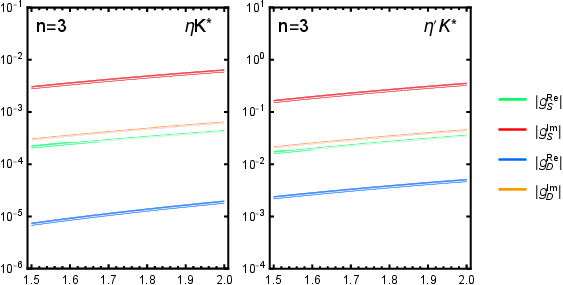}
  \caption{$\alpha$ dependence of $|g_D^{\rm Re}|$, $|g_D^{\rm Im}|$, $|g_S^{\rm Re}|$, and $|g_S^{\rm Im}|$ with $n=3$. The shaded areas result from the errors of the coupling constants.}\label{Fig:etaKstar_n3}
\end{figure}

In the case of $Z_c\to \rho\eta^{(\prime)}$~\cite{Wu:2016ypc} and $Z_b\to \rho\eta^{(\prime)}$~\cite{Wu:2022hck}, the $\rho\eta$ channel is about twice as large as $\rho\eta^\prime$. However, the $\eta K^*$ channel is much lower than $\eta^\prime K^*$ in the case of $Z_{cs}$ decay. In the following, we take $Z_b\to \rho\eta^{(\prime)}$ as an example to illustrate this discrepancy. On one hand, regarding the $Z^{(\prime)}_b$ states as $B^{(*)}\bar{B}^{(*)}$ molecular states, their light hadron decays proceed through charged and neutral charmed meson loops. On the other hand, from Eq.~(\ref{eq:eta-etap mixing}), we can obtain that $\theta_P=-19.1^\circ$ corresponds to the $66\%$ $n\bar n$ component in $\eta$ and the $34\%$ $n\bar n$ component in $\eta^\prime$. In the case of $Z_{cs}\rightarrow \eta^{(\prime)} K^*$, the decay can proceed through two kinds of loops. One kind involves the initial $Z_{cs}$ connecting with the final $K^*$ meson via a strange-charmed meson, and the other involves the final $\eta^{(\prime)}$ coupling to a pair of strange-charmed mesons. The former and the latter loops are labeled by $\mathcal{A}_{D^{(*)}\bar{D}^{(*)}_s D^{(*)}}$ and $\mathcal{A}_{\bar{D}^{(*)}_s D^{(*)} \bar{D}^{(*)}_s}$, corresponding to loops $D^{0} D^{*-}_s D^{*0}$, $D^{*0} D^{-}_s D^{0}$, $D^{*0} D^{-}_s D^{*0}$ and $D^{*-}_s D^0 D^{-}_s$, $D^{*-}_s D^0 D^{*-}_s$, $D^{-}_s D^{*0} D^{*-}_s$ in Table.~\ref{Table:loops}, respectively. The decay amplitudes of $Z_{cs}\rightarrow \eta^{(\prime)} K^*$ are consequently expressed as
\begin{eqnarray}
\mathcal{M}_{Z_{cs}\rightarrow \eta K^*}&=& \frac{\beta}{\sqrt{2}}\mathcal{A}_{D^{(*)}\bar{D}^{(*)}_s D^{(*)}} - \gamma\mathcal{A}_{\bar{D}^{(*)}_s D^{(*)} \bar{D}^{(*)}_s},\nonumber\\
\mathcal{M}_{Z_{cs}\rightarrow \eta^\prime K^*}&=& \frac{\gamma}{\sqrt{2}}\mathcal{A}_{D^{(*)}\bar{D}^{(*)}_s D^{(*)}} + \beta\mathcal{A}_{\bar{D}^{(*)}_s D^{(*)} \bar{D}^{(*)}_s}.
 \label{eq:amplitude_mixing}
\end{eqnarray}
With the mixing angle $\theta_P=-19.1^{\circ}$, the decay amplitudes are $\mathcal{M}_{Z_{cs}\rightarrow \eta K^*}= 0.57\mathcal{A}_{D^{(*)}\bar{D}^{(*)}_s D^{(*)}} - 0.58\mathcal{A}_{\bar{D}^{(*)}_s D^{(*)} \bar{D}^{(*)}_s}$ and $\mathcal{M}_{Z_{cs}\rightarrow \eta^\prime K^*}= 0.41\mathcal{A}_{D^{(*)}\bar{D}^{(*)}_s D^{(*)}} + 0.81\mathcal{A}_{\bar{D}^{(*)}_s D^{(*)} \bar{D}^{(*)}_s}$ with $\mathcal{A}_{D^{(*)}\bar{D}^{(*)}_s D^{(*)}}$ and $\mathcal{A}_{\bar{D}^{(*)}_s D^{(*)} \bar{D}^{(*)}_s}$ being nearly equal. For example, with $\alpha=0.25$, $n=2$, and using the central values of the coupling constants, we have
\begin{eqnarray}
&\mathcal{A}^{\eta K^*}_{D^{(*)}\bar{D}^{(*)}_s D^{(*)}}=[(0.000350-0.005005i)p_{3\beta}p_{2\mu}\nonumber\\
&-(0.002628-0.028282i)g_{\beta\mu}]\epsilon^\mu_1 \epsilon^\beta_2, \\
&\mathcal{A}^{\eta K^*}_{\bar{D}^{(*)}_s D^{(*)} \bar{D}^{(*)}_s}=[(0.000347-0.005141i)p_{3\beta}p_{2\mu}\nonumber\\
&-(0.002698-0.027299i)g_{\beta\mu}]\epsilon^\mu_1 \epsilon^\beta_2, \\
&\mathcal{A}^{\eta^\prime K^*}_{D^{(*)}\bar{D}^{(*)}_s D^{(*)}}=[(0.000386-0.004236i)p_{3\beta}p_{2\mu}\nonumber\\
&-(0.002740-0.028552i)g_{\beta\mu}]\epsilon^\mu_1 \epsilon^\beta_2, \\
&\mathcal{A}^{\eta^\prime K^*}_{\bar{D}^{(*)}_s D^{(*)} \bar{D}^{(*)}_s}=[(0.000378-0.004179i)p_{3\beta}p_{2\mu}\nonumber\\
&-(0.002791-0.027388i)g_{\beta\mu}]\epsilon^\mu_1\epsilon^\beta_2,
\end{eqnarray}
where $p_1 (\epsilon_1)$, $p_2 (\epsilon_2)$ and $p_3$ are the four-momenta (polarization vectors) of the initial state $Z_{cs}$, final states $\mathcal{V}$  and $\mathcal{P}$, respectively. Therefore, the branching ratio of $Z_{cs}\rightarrow \eta K^*$ is highly suppressed relative to that of $Z_{cs}\rightarrow \eta^\prime K^*$. It is expected that $Z_{cs}\rightarrow \eta^\prime K^*$ is SU(3) flavor-allowed, while for $Z_{cs}\rightarrow \eta K^*$, it proceeds via $\eta-\eta^\prime$ mixing induced by SU(3) flavor symmetry breaking. To further verify that the strong suppression of the $\eta K^*$ channel is robust within the present model framework, we rewrite the decay amplitudes $Z_{cs}\rightarrow \eta^{(\prime)} K^*$ in the form,
\begin{eqnarray}
\mathcal{M}_{Z_{cs}\rightarrow \eta^{(\prime)} K^*}&=&[(g_D^{\rm Re}+ig_D^{\rm Im})p_{3\beta}p_{2\mu}\nonumber\\
&&+(g_S^{\rm Re}+ig_S^{\rm Im})g_{\beta\mu}]\epsilon^\mu_1\epsilon^\beta_2,
\label{eq:amplitude_etaK}
\end{eqnarray}
and investigate the dependence of the $g_D^{\rm Re}$, $g_D^{\rm Im}$, $g_S^{\rm Re}$, and $g_S^{\rm Im}$ on the cutoff, the form factor, and the input couplings. From Figs.~\ref{Fig:etaKstar_n2} and \ref{Fig:etaKstar_n3}, the coefficients $g_D^{\rm Re}$, $g_D^{\rm Im}$, $g_S^{\rm Re}$, and $g_S^{\rm Im}$ in the $\eta K^*$ channel are lower than those in the $\eta^\prime K^*$ channel by about two orders of magnitude for both $n=2$ and $n=3$, indicating that the interference cancellation in the $\eta K^*$ channel and the interference enhancement in the $\eta^\prime K^*$ channel are both independent of the form factor. Moreover, these interference and enhancement effects hold over the entire  cutoff range we considered. One can also conclude that the errors of the coupling constants do not change the conclusions drawn here.

From Table.~\ref{Tab:Zcsbr} and Figs.~\ref{Fig:pv1}--\ref{Fig:pv2}, the branching ratios of $Z_{cs}\rightarrow \mathcal{P}\mathcal{V}$ exhibit a certain degree of uncertainty due to $\alpha$. Exploring the $\alpha$ dependence of the ratios among different branching ratios could help further resolve the uncertainties introduced by the inclusion of form factors. We define the ratios of branching ratios as follows
\begin{eqnarray}
	R_{1}&=& \frac{\mathcal{B}[Z_{cs}\rightarrow K\rho]}{\mathcal{B}[Z_{cs}\rightarrow K\omega]} \, ,\nonumber\\
    R_{2}&=& \frac{\mathcal{B}[Z_{cs}\rightarrow K\phi]}{\mathcal{B}[Z_{cs}\rightarrow K\omega]} \, ,\nonumber\\
    R_{3}&=& \frac{\mathcal{B}[Z_{cs}\rightarrow \eta^\prime K^*]}{\mathcal{B}[Z_{cs}\rightarrow \eta K^*]} \, ,\nonumber\\
    R_{4}&=& \frac{\mathcal{B}[Z_{cs}\rightarrow \pi K^*]}{\mathcal{B}[Z_{cs}\rightarrow \eta^\prime K^*]}.
     \label{Eq:ratio}
\end{eqnarray}

\begin{figure}[!htb]
  \centering
 \includegraphics[width=8cm]{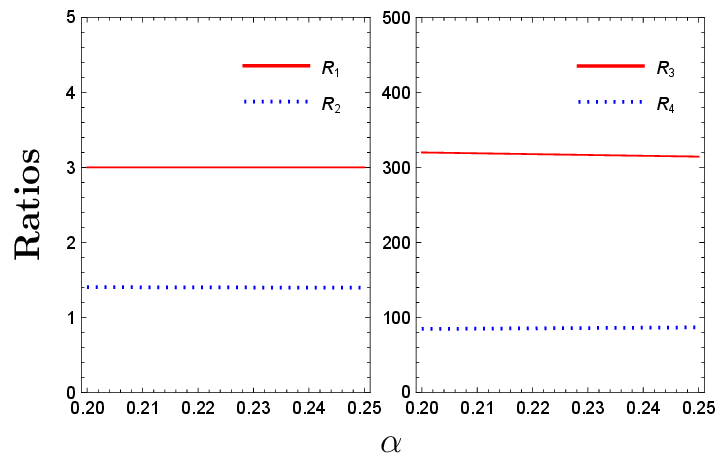}
  \caption{$\alpha$ dependence of the ratios $R_1$, $R_2$, $R_3$, and $R_4$ defined in Eqs.~(\ref{Eq:ratio}) with $n=2$.}\label{Fig:ration2}
\end{figure}

\begin{figure}[!htb]
  \centering
 \includegraphics[width=8cm]{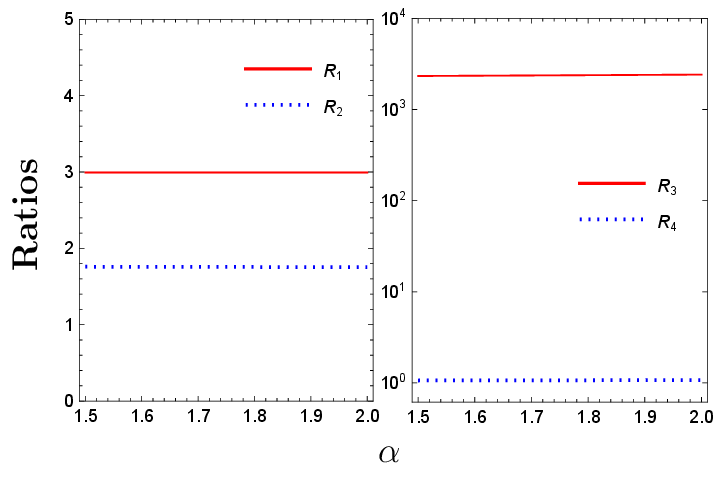}
  \caption{$\alpha$ dependence of the ratios $R_1$, $R_2$, $R_3$, and $R_4$ defined in Eqs.~(\ref{Eq:ratio}) with $n=3$.}\label{Fig:ration3}
\end{figure}

The ratios $R_i$ $(i=1, 2, 3, 4)$ as functions of $\alpha$ are plotted in Fig.~\ref{Fig:ration2} and Fig.~\ref{Fig:ration3} with $n=2$ and $n=3$, respectively. For $n=2$, the ratios are $R_1\simeq3$, $R_2\simeq1.4$, $R_3\simeq315-320$, and $R_4\simeq85-87$ with $\alpha=0.2-0.25$. For $n=3$, they become $R_1\simeq3$, $R_2\simeq1.7$, $R_3\simeq(2.3-2.4)\times10^3$, and $R_4\simeq1.1$ with $\alpha=1.5-2$. The ratios are almost independent of $\alpha$, indicating that the form factor provides a reasonably controlled cutoff for each channel. The ratios $R_1$ and $R_2$ are almost independent of the choice of $n$, whereas $R_3$ and $R_4$ exhibit a strong dependence on $n$. Consequently, $R_1$ and $R_2$ are stable within the form factor scheme adopted in this work, and thus they are relatively robust predictions within the present molecular framework. On the other hand, one may wonder how the $\eta-\eta^\prime$ mixing angle $\theta_P$ affects the decay processes that involve the $\eta^{(\prime)}$ meson, which is of more fundamental significance than the parameter $\alpha$. This motivates us to study the dependence on the mixing angle $\theta_P$.

\begin{figure}
  \centering
 \includegraphics[width=8cm]{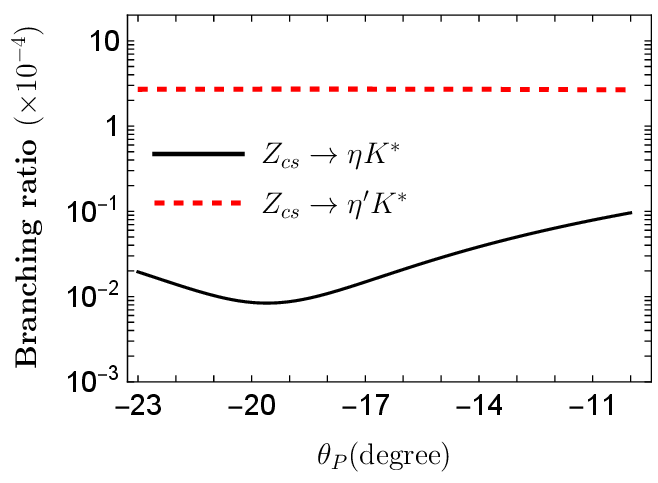}
  \caption{ Branching ratios of $Z_{cs}\to \eta^{(\prime)}K^*$ depending on the mixing angle $\theta_P$ with $\alpha=0.25$.}\label{Fig:brmixing}
\end{figure}

In Fig.~\ref{Fig:brmixing}, we plot the $\theta_P$ dependence of the branching ratios of $Z_{cs}\to \eta^{(\prime)}K^*$ by fixing $\alpha=0.25$. We find that the branching ratio of $Z_{cs}\to \eta^\prime K^*$ is almost independent of $\theta_P$, which is around $2.7\times10^{-4}$. As for $Z_{cs}\to \eta K^*$, the branching ratios
first decrease and then increase with increasing $\theta_P$. To some extent, this behavior reveals how the mixing angle affects our calculated results. Specifically, the decay $Z_{cs}\to \eta K^*$ is more sensitive to the mixing angle $\theta_P$ than $Z_{cs}\to \eta^\prime K^*$. This is easy to understand, since $\mathcal{M}_{Z_{cs}\rightarrow \eta K^*}$ and $\mathcal{M}_{Z_{cs}\rightarrow \eta^\prime K^*}$ exhibit destructive interference and constructive interference, as seen from Eq.~(\ref{eq:amplitude_mixing}), respectively. The branching ratio of $Z_{cs}\rightarrow \eta K^*$ is extremely sensitive to $\theta_P$, and can therefore serve as a probe to determine the $\eta-\eta^\prime$ mixing angle on the theoretical side. However, it is difficult to be implemented on the experimental side, since the branching ratio of $Z_{cs}\rightarrow \eta K^*$ is too small. In contrast, the branching fraction of $Z_{cs}\rightarrow \eta^\prime K^*$ is almost independent of $\theta_P$, making this channel
cleaner with respect to this particular source of uncertainty.

Lastly, we summarize several remarks as follows:
\begin{itemize}
  \item It should be pointed out that our predictions are based on the molecular interpretation of $Z_{cs}(3985)$; other Fock components could alter the numerical results. Owing to SU(3) flavor symmetry, $Z_{cs}(3985)$ and $Z_c(3900)$ exhibit similar decay patterns. As shown in Ref.~\cite{Esposito:2014hsa}, tetraquark models (especially type I) predict that at least one of the $Z_c$ state decays copiously into $\eta_c \rho$, while the molecular model predicts a strong suppression. Investigating the decays of $Z_{cs}(3985)$ under the tretraquark picture may thus yield well-separated predictions between the two scenarios, and render certain decay channels clean discriminants.
  \item Since the final light mesons in the decays considered here are not soft, the calculation cannot be viewed as a controlled chiral expansion, but rather a phenomenological model estimate. Moreover, the effects associated with the use of Eq.~(3) outside the soft-meson regime and with the off-shell extrapolation cannot be reliably quantified within the present framework, and constitute an additional systematic model uncertainty beyond the errors propagated from the input coupling constants.
  \item The absolute branching ratios predicted in this work should be interpreted as order-of-magnitude estimates. Although the branching ratio of each individual channel is rather sensitive to the variation of $\alpha$, the relative branching ratios among the various channels remain fairly stable. We therefore propose that experiments measure the predicted ratios of branching ratio, especially $R_1$ and $R_2$, which would be meaningful for testing the molecular interpretation of $Z_{cs}$.
\end{itemize}

\section{SUMMARY}
\label{sec:summary}

The discovery of the strange hidden-charm tetraquark states $Z_{cs}(3985)$, $Z_{cs}(4000)$, and $Z_{cs}(4220)$ has opened new avenues for understanding multiquark dynamics. While previous studies focused on open-charm and hidden-charm decays, this work systematically investigates the charmless decays $Z_{cs} \to \mathcal{P}\mathcal{V}$ within a molecular picture. We adopt the effective Lagrangian approach and evaluate triangle-loop diagrams, with form factors introduced to regularize the ultraviolet behavior. By reproducing the experimental upper limit on the branching ratio of $Z_{c}(3900)\rightarrow \pi\omega$, we constrain the cutoff parameter and evaluate the branching ratios of $Z_{cs} \to \mathcal{P}\mathcal{V}$ with both the dipole and tripole form factors. Numerical results from dipole form factor show that the branching ratios for $Z_{cs} \to \pi K^*$ can reach the order of $10^{-2}$, which is of the same order of magnitude as the hidden charm decay modes and may be detectable in future measurements. The branching ratios for $Z_{cs} \to K\rho/\omega/\phi$ and $\eta^\prime K^{*}$ are of the order of $10^{-5}-10^{-4}$, two orders of magnitude lower than those of the hidden charm decay modes. For the $\eta K^{*}$ channel, the branching ratio is of the order of $10^{-7}$, which is highly suppressed. The branching ratios from the tripole form factor are not consistent with those from the dipole form factor, indicating a certain degree of model dependence in the present work.

Furthermore, we analyze the dependence of these decay rates on the model parameter $\alpha$ and the $\eta$-$\eta'$ mixing angle $\theta_P$. The estimated branching fraction ratios for various decay channels exhibit little dependence on the cutoff parameter. The ratios $\frac{\mathcal{B}[Z_{cs}\rightarrow K\rho]}{\mathcal{B}[Z_{cs}\rightarrow K\omega]}$ and $\frac{\mathcal{B}[Z_{cs}\rightarrow K\phi]}{\mathcal{B}[Z_{cs}\rightarrow K\omega]}$ are found to be nearly independent of the form factor, making them relatively robust predictions within the present molecular
framework. However, ratios $\frac{\mathcal{B}[Z_{cs}\rightarrow \eta^\prime K^*]}{\mathcal{B}[Z_{cs}\rightarrow \eta K^*]}$ and $\frac{\mathcal{B}[Z_{cs}\rightarrow \pi K^*]}{\mathcal{B}[Z_{cs}\rightarrow \eta^\prime K^*]}$ are highly dependent on the choice of the dipole or tripole form factor.

We also estimate the branching ratios of $Z_{cs}\to \eta^{(\prime)}K^*$ depending on the mixing angle $\theta_P$ by fixing $\alpha=0.25$. The branching fraction of $Z_{cs}\rightarrow \eta K^*$ is extremely sensitive to $\theta_P$, making it a potential probe for the $\eta$-$\eta^\prime$ mixing angle, while that of $Z_{cs}\rightarrow \eta^\prime K^*$ is nearly independent of $\theta_P$, making this channel cleaner with respect to this particular source of uncertainty. Future experimental measurements of these charmless decay channels, particularly at BESIII, LHCb, and Belle, can help constrain the $\eta$-$\eta'$ mixing angle and further clarify the internal structure of the $Z_{cs}$ states.

\section*{ACKNOWLEDGMENTS}

This work is supported by the National Natural Science Foundation of China under the Grants No. 12405093, No. 12175037, No. 12335001, No. 12475081, and No. 12475197, as well as supported, in part, by the National Key Research and Development Program under Grants No. 2024YFA1610503 and No. 2024YFA1610504. It is also supported by the Taishan Scholar Project of Shandong Province under Grant No. tsqn202607074, the Natural Science Foundation of Shandong Province under Grants No. ZR2025MS04 and No. ZR2022ZD26, and the Natural Science Foundation of Henan Province under Grant No. 262300421060.

\section*{DATA AVAILABILITY}

There are no publicly available research data or software supporting this manuscript. Requests for further information or data should be sent to the authors.

\bibliographystyle{unsrt}
\bibliography{references}

\end{CJK}	
\end{document}